\documentclass[12pt]{article}

\def\fnote#1#2{\begingroup\def\thefootnote{#1}\footnote{#2}\addtocounter{footnote}{-1}\endgroup}

\usepackage{amssymb}
\usepackage{makeidx}
\usepackage{epsf}

\makeindex

\def\inbar{\vrule height1.5ex width.4pt depth0pt}
\def\IB{\relax{\rm I\kern-.18em B}}
\def\IC{\relax\,\hbox{$\inbar\kern-.3em{\rm C}$}}
\def\ID{\relax{\rm I\kern-.18em D}}
\def\IE{\relax{\rm I\kern-.18em E}}
\def\IF{\relax{\rm I\kern-.18em F}}
\def\IG{\relax\,\hbox{$\inbar\kern-.3em{\rm G}$}}
\def\IH{\relax{\rm I\kern-.18em H}}
\def\II{\relax{\rm I\kern-.18em I}}
\def\IK{\relax{\rm I\kern-.18em K}}
\def\IL{\relax{\rm I\kern-.18em L}}
\def\IM{\relax{\rm I\kern-.18em M}}
\def\IN{\relax{\rm I\kern-.18em N}}
\def\IO{\relax\,\hbox{$\inbar\kern-.3em{\rm O}$}}
\def\IP{\relax{\rm I\kern-.18em P}}
\def\IQ{\relax\,\hbox{$\inbar\kern-.3em{\rm Q}$}}
\def\IR{\relax{\rm I\kern-.18em R}}
\def\IT{\relax{\rm I\kern-.18em T}}
\def\ZZ{\relax{\sf Z\kern-.4em Z}}

\def\a{\alpha}   \def\b{\beta}    \def\g{\gamma}  \def\d{\delta}
\def\e{\epsilon} \def\G{\Gamma}     \def\l{\lambda}
    \def\Om{\Omega} \def\si{\sigma}
\def\Si{\Sigma}

 \def\cH{{\cal H}} \def\cI{{\cal I}}
  
 \def\cN{{\cal N}} \def\cO{{\cal O}}
\def\cP{{\cal P}}  
  \def\cU{{\cal U}}

 \def\Gwhat{{\widehat{G}}}

\def\rmA{{\rm A}}  \def\rmB{{\rm B}}  
\def\rmD{{\rm D}}    
    
    \def\rmN{{\rm N}}

    \def\rmAutF{{\rm AutF}}  
\def\rmBP{{\rm BP}}

\def\rmGal{{\rm Gal}}    \def\rmGL{{\rm GL}}     \def\rmGSO{{\rm GSO}}

       \def\rmHom{{\rm Hom}}

\def\rmND{{\rm ND}}

\def\rmSO{{\rm SO}}          
  \def\rmSL{{\rm SL}}      

\def\rmWD{{\rm WD}}

        \def\rmaff{{\rm aff}}    \def\rmalg{{\rm alg}}
     \def\rmcoh{{\rm coh}} 

  \def\rmdeg{{\rm deg}}   \def\rmdet{{\rm det}}      \def\rmdiag{{\rm diag}}
\def\rmdim{{\rm dim}}
           
\def\rmhol{{\rm hol}}   
   
 \def\rmlcm{{\rm lcm}}         
\def\rmrk{{\rm rk}}
   \def\rmmod{{\rm mod}}   \def\rmmon{{\rm mon}}

\def\rmth{{\rm th}}      \def\rmtr{{\rm tr}}

\def\afrak{{\mathfrak a}}  \def\cfrak{{\mathfrak c}}    
  \def\gfrak{{\mathfrak g}}
 \def\mfrak{{\mathfrak m}}
\def\nfrak{{\mathfrak n}} 
\def\pfrak{{\mathfrak p}}

   \def\mathC{{\mathbb C}}     
 \def\mathF{{\mathbb F}}   \def\mathG{{\mathbb G}}      
   \def\mathN{{\mathbb N}}
 \def\mathP{{\mathbb P}}   \def\mathQ{{\mathbb Q}} \def\mathR{{\mathbb R}}
\def\mathZ{{\mathbb Z}}

\def\fnote#1#2{\begingroup\def\thefootnote{#1}\footnote{#2}\addtocounter{footnote}{-1}\endgroup}

\def\beq{\begin{equation}}
\def\eeq{\end{equation}}
\def\bea{\begin{eqnarray}}
\def\eea{\end{eqnarray}}

\def\lleq#1{\label{#1}\eeq}
\let\nn=\nonumber

\def\notin{\ \hbox{{$\in$}\kern-.51em\hbox{/}}}

\def\lra{\longrightarrow}
\def\lolra{{\longleftrightarrow}}

  \def\E1Fq{E_1/\IF_q}

\def\notdiv{{\relax{~|\kern-.34em /~}}}

\def\oa{{\overline{a}}} 
   \def\ou{{\overline{u}}}   \def\oz{{\overline{z}}}

  \def\omathQ{{\overline{\mathQ}}}

    \def\osi{{\overline{\si}}}  

\def\boxit#1{
\vbox{\hrule height1pt\hbox{\vrule width1pt\kern0.3cm
\vbox{\kern0.3cm\hbox{$\displaystyle#1$}\kern0.3cm}\kern0.3cm\vrule
width1pt}\hrule height1pt}}

\begin{document}

\hfill \phantom{  \hfill \today ~}

\vskip 1.2truein

\centerline{\large {\bf String automorphic motives of nondiagonal varieties}} 

\vskip .4truein

\centerline{\sc Rolf Schimmrigk\fnote{$\diamond$}{netahu@yahoo.com, rschimmr@iusb.edu}}

\vskip .3truein

\centerline{Indiana University South Bend}

\centerline{1700 Mishawaka Ave., South Bend, IN 46634}

\vskip 1.1truein
\baselineskip=18pt

\centerline{\bf Abstract} 
 \begin{quote}
 In this paper automorphic motives are constructed and analyzed with a view toward the 
 understanding of the geometry of compactification manifolds in string theory in terms of 
 the modular structure of the worldsheet theory. The results described generalize a 
 framework considered previously in two ways, first by relaxing the restriction to modular 
 forms, and second by extending the construction of motives from diagonal varieties to 
 nondiagonal spaces. The framework of automorphic forms and representations is described 
 with a view toward applications, emphasizing the explicit structure of these objects. 
 \end{quote}

\vskip .3truein

\renewcommand\thepage{}
\newpage

 \parindent=0pt

 \pagenumbering{arabic}
 \baselineskip=15pt

 \tableofcontents

\vfill \eject

\parskip=.16truein 
\baselineskip=21.3pt 
\parindent=0pt

\section{Introduction}

The purpose of the present paper is to extend and generalize previous results on the construction 
 of  the compact "internal" varieties in string theory  from first principles by using 
 the  worldsheet theory. The basic idea of this program is to associate pure or mixed motives 
 to conformal field theoretic modular forms and to use these motives as the geometric building 
blocks of the geometry spanned the extra dimensions predicted by string theory. 
 Gluing these motives together then determines 
the global structure of the varieties. Previous work in this direction was restricted in two ways. 
First, the class of varieties was restricted to diagonal hypersurfaces, with
  conformal field theories given by exactly solvable diagonal  Gepner models. 
 Within this class of spaces it is possible to consider deformation families that correspond 
 to deformation of the conformal field theory along marginal operators.  It can be shown 
 that certain types of singular fibers in such deformation families  that 
   lead to phases that are modular, with forms that can be constructed from the $N=2$ 
 supersymmetric minimal factors. Second, the relation between the structure of the conformal 
 field theory $T_\Si$ on the  worldsheet $\Si$ and the motives $M(X)$ of  the compact variety $X$ 
 obtained was based purely on the identification of the motivic $L$-function $L(M,s)$ 
 with $L$-functions $L(f_i,s)$ associated to modular cusp forms $f_i$ that are derived from 
 the theory $\Si$.   What is needed is a more conceptual picture  that explains the
experimental results based on $L$-function identities alone.

 The main motivation of the work described here is to relax the condition of diagonality to 
 admit configurations for which the moduli space does not necessarily contain diagonal 
 fibers, and to relax the constraint of modularity.  This establishes that the framework 
 proposed for an emergent spacetime program via automorphic motives extends to 
 more general spacetimes. 
 An extension of modularity is provided by the more general framework  
of automorphic forms $\phi$, and their associated representations $\pi$, as considered 
by Harish-Chandra \cite{hc59, hc68}, Langlands \cite{rl70, rl79}, and many other mathematicians,
 in combination with the conjecture that all (pure) motives are automorphic. 
While in essence the relation between motives and automorphic forms is again mediated by 
the $L$-function, $L(\pi,s)=L(M,s)$, the structure of these automorphic objects allows to make the 
expected relation more precise. As a first step in this direction the automorphic nature of some 
higher rank motives is analyzed. The strategy followed in the present work is a generalization of 
the one considered first in \cite{rs08}, where the automorphy of K3 fibered diagonal 
Calabi-Yau varieties 
was established by constructing the varieties via the twist map introduced in \cite{hs95,hs99}
  (see also the independent work of Voisin \cite{v93} and later work by Borcea \cite{b97}). 
 Here this approach will be applied to motives of nondiagonal spaces.

This paper is organized to start with the geometry as encoded in motives, then continues 
to the number theoretic structure of these motives, and then proceeds to automorphic forms 
 and representations. In the final part these ingredients are combined in the discussion of some 
 examples. More precisely, the structure is as follows. Section 2 introduces the varieties considered 
 in this work and discusses their arithmetic structure. Section 3 outlines the concept of motives, while 
 Section 4 describes the construction of the motives relevant for nondiagonal varieties.  Section 5 
 introduces some necessary number theory that provides a link between the motives and automorphic 
 forms, while Section 6 describes the conceptual structure of automorphic forms necessary for understanding 
the link to motives in a conceptual way. Section 7 describes how these two different types of objects 
can be related in a transparent way with the least amount of machinery and Section 8 makes the general 
framework concrete for the class of $\Om$-motives. In Sections 9 and 10 applications are given in the 
context of modular and automorphic motives for nondiagonal spacetimes.

\section{Arithmetic of diagonal and nondiagonal varieties}

The most important quantity associated to a motive is its $L$-function, an object obtained via the 
 local zeta functions of Artin and Schmidt
 \beq
  Z(X/\mathF_q,t) ~=~ \exp\left(\sum_{r\geq 1} \frac{N_{q,r}(X)}{r}t^r\right), 
 \eeq
 where $N_{q,r}(X)$ counts the number of points $\#(X/\mathF_{q^r})$ of the variety $X$ over the 
  finite field $\mathF_{q^r}$. For curves and simple types of varieties it is easily seen that the zeta 
 function decomposes into factors that are associated to the cohomology groups $H^j(X)$ of 
 $X$, and it was conjectured by Weil \cite{w49} that this is the case more generally.  
This cohomological 
 interpretation of the Artin-Schmidt zeta functions
   \beq
 Z(X_n/\mathF_p,t)
   = \frac{\prod_{j=1}^n \cP_p^{2j-1}(X_n,t)}{\prod_{j=0}^n \cP_p^{2j}(X_n,t)},
 \eeq
 proven by Grothendieck \cite{g65}, shows that only a finite amount of information has to be 
computed in order to determine this object. 
Here $\cP_p^0(X_n,t) = 1-t$, $\cP_p^{2n}(X_n,t) = 1-p^nt$ and the remaining 
 $\cP_p^i(X_n,t)$ are polynomials whose degree is 
given by the Betti numbers $b^i = \rmdim_\mathR H^i(X_n)$ of the variety.
The global Hasse-Weil $L$-function is obtained by combining the local factors $\cP_p^j(X,t)$
 as 
 \beq
 L(H^j(X),s) ~ \doteq~ \prod_p \frac{1}{\cP_p^j(X,p^{-s})}.
 \eeq
   Details about bad primes will not be of relevance in this paper.

For the case of $\Om$-motives of hypersurfaces of weighted projective spaces this arithmetic 
 structure can easily be computed directly, if very inefficiently. A more conceptual analysis of 
the cardinalities $N_{q,r}$  is useful not only for calculational efficiency, but it 
is also necessary in order to extract from the cohomological zeta functions
 the local factors of the motives that are obtained via projectors acting on the cohomology.
 Both of these issues can be addressed as follows.
 
Consider a hypersurface $X_n^d$ of dimension $n$ and degree $d$ embedded in weighted 
 projective spaces $\mathP_{(w_0,...,w_{n+1})}$ with weight vectors denoted by 
 $(w_0,...,w_{n+1}) \in \mathN^{n+2}$.  For smooth hypersurfaces of degree $d$ the 
 monomial part $H_\rmmon(X_n^d)$ of the cohomology of $X_n^d$ 
 is isomorphic to a subset $\cU_\rmcoh$ of integral vectors defined as 
 \beq
 \cU_{n,d} ~=~ \left\{u\in \mathZ^{n+2}~{\Big |}~ 0\leq u_i \leq d_i-1, ~d|\sum_i u_i w_i\right\},
 \eeq
i.e. $H_\rmmon(X_n^d) ~\cong ~ \cU_\rmcoh \subset \cU_{n,d}$.

 Associated to the elements of $\cU_{n,d}$ are products of Gauss sums, 
 defined in terms of two characters associated to finite fields $\mathF_p$. The first 
 is a character on the multiplicative group 
 $\mathF_p^\times$ of $\mathF_p$ into the cyclic group $\mu_{p-1}$
 generated by the $(p-1)^{\rm st}$ root of unity
 \beq
  \chi_p: ~~\mathF_p^\times ~\lra ~\mu_{p-1},
 \eeq
 defined by $\chi_p(v) = \xi_{p-1}^m$, where $\xi_n = e^{2\pi i/n}$, and the integer $m$ is determined by 
 the generator $g\in \mathF_p^\times$ as $v=g^m$. The second is an additive character 
 \beq
 \Psi_p: ~~\mathF_p ~\lra ~ \mu_p
 \eeq
 can be defined as $\Psi_p(v) = \xi_p^v$. With these ingredients the Gauss sums, defined as
   \beq
  G_{n,p} ~=~ \sum_{u\in \mathF_p^\times} ~\chi_p^n(u) \psi_p(u),
 \eeq
 can be combined into the Gauss sum products defined as
 products 
 \beq
 \mathG^{(n)}_p(u) ~=~ \prod_{i=0}^{n+1} G_{-u_iw_ik,p},
 \eeq
  were $k=(p-1)/d \in \mathZ$. These objects can be generalized to 
 finite extension $\mathF_q$ of $\mathF_p$ with $q=p^r$ for some integer $r\in \mathN$ 
 via the trace operator from $\mathF_q$ to $\mathF_p$ (more details can be found e.g. in \cite{kls10}). 

Cardinality problems of algebraic varieties are most often formulated in the $p$-padic framework, 
following the work of Dwork in the 1960s. For the understanding of the nondiagonal 
 motives constructed below it is of advantage to consider the complex framework instead. 
 A computation outlined in \cite{kls10}  shows that the multiplicative affine cardinalities
  $N_p^\times(X_n^d)_\rmaff$ of Brieskorn-Pham hypersurfaces of arbitrary type 
 \beq
 X_n^{d,\rmBP} ~=~ \left\{ \sum_{i=0}^{n+1} z_i^{d_i} ~=~ 0 \right\} ~\subset ~ 
   \mathP_{(w_0,...,w_{n+1})} 
 \lleq{diag-vars}
  is given by 
  \beq
 N_p^\times(X_n^{d,\rmBP})_\rmaff 
   ~=~ \frac{(p-1)^{n+2}}{p} ~+~ \frac{(p-1)}{p} \sum_{u_i\in \cU_{n,d}} \mathG^{(n)}_p(u).
 \lleq{bp-cards}

An extension of this result can be obtained with similar methods as in \cite{kls10} for varieties
 that are nondiagonal.
 In the present paper the focus is on weighted hypersurfaces of the type 
 \beq
  X_n^{d,\rmND} ~=~ \left\{ \sum_{i=0}^n z_i^{d_i} ~+~ z_n z_{n+1}^{d_{n+1}} ~=~ 0\right\} 
   \subset \mathP_{(w_0,....,w_{n+1})}.
 \lleq{1-tadpoles}
 These spaces are of interest because they contain the class of all nondiagonal Gepner models \cite{g87}
 (constructed in  \cite{ls89, fkss89})
 and they overlap with the more general class of Kazama-Suzuki models.

 The formula that gives the cardinalities for these spaces can be written in the form 
 \beq  
  N_p^\times(X_n^{d,\rmND}) 
   ~=~  \frac{(p-1)^{n+2}}{p} ~+~ \frac{(p-1)}{p} \sum_{u_i\in \cU_{n,d}} 
               \mathG_p^{(n-2)}(u)
        G_{-w_i(u_nd_{n+1}-u_{n+1})\frac{k}{d_{n+1}},p} G_{-u_{n+1}\frac{kd}{d_{n+1}},p}.\nn \\
 \lleq{nd-cards}
 These multiplicative affine cardinalities can be used to compute the projective motivic cardinalities 
  that play a central role in the present work.

\section{Motives}

Motives can be thought of in two different ways, either in terms of Grothendieck's formulation 
involving correspondences,  or as Galois representations.
 A discussion of both the original Grothendieck notion as well as the Galois theoretic 
 picture of these objects in a physical context can be found in \cite{rs08}. An in-depth discussion 
of many aspects of motives can be found in the illuminating collection of articles in ref. \cite{jks94}.
 In this section the general structure of motives is briefly described  while in the next section the 
motives relevant for the present work are constructed. 

\subsection{Cohomological realizations of motives}

The motives of interest in the present work can be thought of as realized by subspaces of the 
intermediate cohomology  of a variety $X_n$ of complex dimension $n$
 \beq
  H(M) ~\subset ~ H^n(X_n).
 \eeq
 For such motives their weight $w_M$, defined as the degree of the cohomology
  $w_M=\rmdeg~H(M)$,  is given by the dimension of 
the variety $w_M= \rmdim_\mathC X_n$. 
 More precisely, the Hodge decomposition of the variety induces a decomposition 
of the realization $H(M)$ as 
 \beq
   \bigoplus_{\stackrel{i=1}{r_i+s_i=w_M}}^{\rmrk(M)} H^{r_i,s_i}(M) ~\subset ~H^n(X_n),
 \eeq
 where $\rmrk(M)$ is the rank of the motive. The Hodge decomposition admits an action of the 
multiplicative group $\mathC^\times$ via the characters $\chi^{r,s}$ defined by
 \beq
 \chi^{r,s}(z) ~=~ z^{-r} \oz^{-s}.
 \eeq
 
\subsection{Tensor product of motives}

One of the fundamental reasons why motives are more useful that monolithic varieties in the context 
of the emergent spacetime program via automorphic forms and representations is that they can be 
viewed as the simplest building blocks which can be used to build more complicated structures. One 
of the constructions that facilitate this process is the tensor product. This product is reminiscent of 
the tensor product of conformal field theories, motivating the picture that perhaps both types of 
objects form structured sets (categories) that eventually might be shown to be isomorphic. 
 More concretely
 tensor products are useful because when present they allow to associate lower rank automorphic 
objects to higher rank irreducible motives. For pure motives $M_i$ of rank $\rmrk(M_i)$ and weights 
 $w_M(M_i)$ corresponding rank and weight of the tensor product are given by
 \bea
  \rmrk~M_1\otimes M_2  &=& \rmrk~M_1 \cdot \rmrk~M_2 \nn \\
    w_M(M_1\otimes M_2) &=& w_M(M_1) + w_M(M_2).
 \eea

\subsection{$L$-functions of motives}

The concrete functional relation between the worldsheet theory $T_\Si$ and the variety $X$ is 
 provided by an identity of the $L$-functions that are associated to the different kinds 
 of objects in these models.  As noted above, $L$-functions of a  variety $X$ are obtained 
 via the Weil-Grothendieck factorization of the Artin-Schmidt zeta functions
  as  cohomological functions that  encode the information obtained by probing a variety 
with varying, but finite, resolutions that are given by the size of the primes $p \in \mathN$ that 
 determine the order of the finite fields $\mathF_p$. By using Grothendieck's notion of a motive $M$
 as an object defined by a correspondence, one can think of its cohomological realization $H(M)$ as 
 given by a projector acting on the cohomology groups of the variety.  
 The resulting three-fold factorization allows to construct the global $L$-function of a motive $M$
 as the product
 \beq
  L(M,s) ~=~ \prod_p L_p(M,s),
 \eeq
 where the local factors $L_p(M,s) = \cP_p(M,p^{-s})^{-1}$ are given for motives $M$ with 
 $H(M) \subset H^n(X_n)$ for varieties of dimension $n$ by 
 \beq
 \cP_p(M,t) ~=~ \rmdet\left({\bf 1} + H_p(M) t\right).
 \eeq
 Here $t$ is a formal variable, which is set to $t=p^{-s}$, $s\in \mathC$, and 
 $H_p(M)$ is a matrix whose rank is equal to the rank of the motive,  computable
 in the context of weighted hypersurfaces in terms of Gauss sums. 
 This will be made more explicit further below.
 One of the key issues in the theory of motivic 
$L$-functions is to establish  continuation to the whole complex plane.

\section{$\Om$-motives}

\subsection{General concept of the $\Om$-motive}

The general notion of the $\Om$-motive is based on the idea to associate a number field $K_X$ to 
the compact variety and to define the motive by orbit of the Galois group $\rmGal(K_X/\mathQ)$
 of $K_X$ over $\mathQ$ on a distinguished cohomology class $\Om$ \cite{rs08}. This concept 
applies to all Calabi-Yau varieties, and more generally to Fano varieties of special type. This section 
briefly outlines the general framework before applying it to diagonal and nondiagonal varieties.

 The conceptual basis of the field $K_X$ is motivated by the Weil conjectures for varieties restricted to finite 
 fields $\mathF_p$, for any finite prime $p$. Denote this restriction for varieties of complex dimension $n$ 
 by $X_n/\mathF_p$.
 The Weil-Grothendieck factorization of the local zeta functions leads to the 
  natural question of what the 
nature is of the complex numbers that arise in the factorization these polynomials   
  \beq
  \cP_p^i(t) ~=~ \prod_j (1+\delta^i_j(p)t).
  \eeq 
 For special classes of manifolds it is easy to see that the 
  $\delta_j^i(p)$ are algebraic numbers, and it was 
 Weil's supposition that this always the case and that for smooth manifolds the norm of these numbers 
 is determined by the primes $p$ as 
  \beq
  |\delta_j^i(p)|= p^{i/2}.
  \eeq 
 This conjecture was one of the main motivations for Grothendieck's 
 development of arithmetic algebraic geometry before Deligne's proof using Grothendieck's framework
  \cite{d74}. 

Given the algebraic nature of the coefficients $\d_j^i(p)$ provided by the 
proof of the Weil conjectures as given by Grothendieck and Deligne it makes sense to define a number field 
 $K_{X_n}$ associated to an algebraic variety $X_n$ of dimension $n$ as 
   \beq
    K_{X_n} := \mathQ(\{\d^n_j(p)\}_{\forall j}).
  \eeq
  The $\Om$-motive for any Calabi-Yau variety (CY) or Fano 
variety of special type (SF) is then defined as the orbits of the $\Om$-form under the 
Galois group $\rmGal(K_X/\mathQ)$ of the field $K_X$ 
 \beq
  H(M_\Om) ~=~ \left\langle \rmGal(K_X/\mathQ), \Om\right\rangle.
 \eeq
 For Calabi-Yau $n$-folds $X_n$ one has $\Om\in H^{n,0}(X_n)$, while for Fano varieties of special type 
  $\Om \in H^{n+Q-1,Q-1}(X_n)$. The Fano varieties specialize to
 Calabi-Yau spaces for $Q=1$ (more details of these varieties are discussed below).

\subsection{$\Om$-motives of diagonal and nondiagonal hypersurfaces}

The concept of the $\Om$-motive given above can be made more concrete and computable 
 for hypersurfaces in toric or weighted projective spaces. 
 For diagonal weighted Calabi-Yau hypersurfaces $X_n^d$ of complex dimension $n$ and degree $d$
 it reduces to the motive defined as the Galois orbit of the holomorphic $(n,0)$-form with 
 respect to the Galois group $\rmGal(\mathQ(\mu_d)/\mathQ)$ defined by the cyclotomic field 
 $K_X=\mathQ(\mu_d)$, where $\mu_d = \langle \xi_d\rangle $ is the cyclic group generated 
 by $\xi_d=e^{2\pi i/d}$.  The degree of such varieties is given by the lowest common multiple 
 of the degrees $d_i$ of the monomials $z_i^{d_i}$ and it would seem natural to adopt the 
same definition for nondiagonal hypersurfaces. However, while the orbits so constructed can be 
 used to extract the motive, they are not  irreducible. To obtain irreducible motives for 
  varieties of type (\ref{1-tadpoles}) define an integer  $v\in \mathZ$ as
  the lowest common multiple of the degrees $d_i$ of the variables $z_i$ for $i\neq n$ 
  of the  monomials of the defining polynomial 
 \beq
 v ~=~ \rmlcm~\{d_i\}_{i\neq n}.
 \eeq

 The field $K_X$ associated to the hypersurface $X$ is then the abelian field $K_X = \mathQ(\mu_v)$ 
 and the motive is given by the  orbit of the $\Om$ form, which now can be written in a more explicit 
form  because of the lattice representation of the monomial cohomology given by $\cU_{n,d}$ in Section 2.
 In this realization of the differential form $\Om$ for both CYs and SFs 
 corresponds to the unit vector  $u_\Om:=(1,...,1) \in \mathZ^{n+2}$ for a variety of complex dimension $n$.
 The explicit form of the $\Om$-form makes it possible to define the 
 action of the Galois group $\rmGal(K_X/\mathQ)$ more concretely. 
 The precise form of this action is different in the diagonal and the nondiagonal cases: 
 while in the diagonal case the modding is by $d_i$ in all the components
 this does not hold for nondiagonal hypersurfaces. In this case the form of the action  
is determined by the same modding condition for all components except the variable that enters linear 
 in the coupling 
 \beq
  \si_r(u_{\Om,i}) ~\equiv ~ru_{\Om,i}(\rmmod~d_i),~~~~~~i\neq n,
 \eeq
 while the $n^\rmth$ component is determined by the constraint $d|\sum_i w_iu_i$, 
leading to a unique result for $\si_r(u_\Om)$. 
 In the following the  resulting  orbit of vectors obtained from the Galois group is denoted by 
 \beq
 \cU_\Om ~=~ \left\{\si_r(u_\Om) ~{\Big |}~\si_r \in \rmGal(K_X/\mathQ) \right\} ~
   \subset ~ \cU_\rmcoh \subset \cU_{n,d}.
 \eeq
 The cohomological representation of the $\Om$-motive can now be written as 
 \beq
  H(M_\Om) ~\cong ~ \left\langle \rmGal(\mathQ(\mu_v)/\mathQ), u_\Om\right\rangle,
 \eeq 
 where $u_\Om=(1,1,\dots,1) \in \cU_{n,d}$.  The $\Om$-motives of both diagonal and nondiagonal 
varieties (\ref{1-tadpoles}) are of CM type, as implied by the cardinality formulae
 (\ref{bp-cards})  and (\ref{nd-cards}).

\subsection{Cardinalities and $L$-functions of $\Om$-motives}

The projective motivic cardinalities can be obtained by considering the Galois orbit of vectors 
 $\cU_\Om \subset \cU$ spanned by the vector  $u_\Om$. Viewing this orbit as a realization of 
the $\Om$-motive $H(M_\Om) = \langle \rmGal(K_X/\mathQ),u_\Om\rangle$,
 as described above,
 the motivic cardinalities can be obtained from the cardinality formula given above 
as
 \beq
  N_p(M_\Om) ~=~ \frac{1}{p} \sum_{u\in \cU_\Om} \mathG_p(u).
 \eeq
 There are different conventions for the local polynomials $\cP_p(M,t)$ of the motivic 
 part of the zeta functions. In the present paper these factors are  written as 
  $\cP_p(M_\Om,t) = \rmdet \left({\bf 1} + H_p(M_\Om) t\right)$, following the notation 
 used earlier. With this notation the matrices $H_p(M_\Om)$ can be expressed in terms 
 of the Gauss sum products of the Galois orbit as
 \beq
 H_p(M_\Om) ~=~ (-1)^{w+1}\frac{1}{p}\left(\matrix{
         \mathG_p(\si_1(u_\Om)) &             & \cr
                                                &\ddots  & \cr
                &   &\mathG_p(\si_r(u_\Om))\cr}\right),
 \eeq
 leads with $\rmtr ~H_p(M_\Om) = (-1)^{w+1}N_p(M_\Om)$ to the $L$-function coefficients. 

Defining 
 the local factors of the $L$-function at the good primes essentially as the inverse of the polynomials 
 $\cP_p(M_\Om,t)$ leads to  
  \beq
  L(M_\Om,s) ~=~ \prod_p \frac{1}{\cP(M_\Om,p^{-s})},
 \eeq
 hence the coefficients of the expansion 
 \beq 
 L(M_\Om,s) ~=~ \sum_n \frac{a_n(M_\Om)}{n^s} 
 \eeq
 are determined as
 \beq
  a_p(M_\Om) ~=~ - \rmtr~H_p(M_\Om) ~=~ (-1)^w N_p(M_\Om).
 \eeq

With these local factors the $L$-functions can be computed explicitly and the comparison with 
automorphic $L$-functions $L(\phi,s)$ can be attempted. This is the strategy followed
 in the automorphic spacetime program in the context of diagonal varieties in \cite{rs08} 
and references therein. 

\section{Algebraic Hecke characters}

Algebraic Hecke characters provide the key ingredients  that allow to link  
 motives with complex multiplication to automorphic objects. 
They are also useful 
because they provide the simplest context in which algebraicity can be made explicit.

\subsection{Infinity type}

To gain a more conceptual picture it is useful to note that up to factors of $p$ the
 Gauss sum products $\mathG_p$ are essentially Jacobi sums and therefore 
 define algebraic Hecke characters $\Psi$, as first observed by in ref. \cite{w52}.  Weil's notion 
 of Hecke characters of type $A_0$ has been reformulated by Deligne \cite{d77} 
 into the notion of an algebraic Hecke character as follows. Hecke characters  
 can be associated to arbitary number fields $K$ and are defined as maps on the group 
  $\cI_\mfrak(\cO_K)$ of fractional ideals prime to the modulus ideal $\mfrak$.  
  They are characterized by their behavior on the principal ideals $\afrak=(z)$, $z\in K$
 where the algebraic  characters $\chi_\rmalg: ~K^\times  \lra E$ are defined by elements 
of the group ring
 \beq
  S ~=~ \sum_{\ell}  n^\ell \si_\ell ~\in~ \mathZ[\rmHom(K,\omathQ)]
  \eeq
 as
 \beq
  \chi_\rmalg(z) ~=~ z^{S} ~=~ \prod_{\ell} \si_\ell(z)^{n^\ell}.
 \eeq

{\bf Definition.}
  ~Let $K$ be a number field and $\cO_K$ its ring of integers. For $\cfrak \subset \cO_K$
 an integral ideal denote by $\cI_\cfrak(K)$ the group of fractional ideals of $K$ that are prime to 
 $\cfrak$ and by $\cI_\cfrak^p(K)$ the principal ideals $(z)$ of $K$ such that $z\equiv 1(\rmmod~\cfrak)$.
 An algebraic Hecke character $\Psi$ modulo $\cfrak$ is a multiplicative function $\chi$ on $\cI_\cfrak(K)$ 
 whose structure on $(z) \in \cI_\cfrak^p(K)$ is determined by an algebraic character $\chi_\rmalg$
 \beq
  \Psi((z)) ~=~  \chi_\rmalg(z) ~=~ z^S
 \eeq
 in terms of the infinity type $S$. The integer $w=n_\ell+ n_{c\ell}$ for all $\ell$, 
 with $\si_{c\ell} = \osi_\ell$, is called the weight of the character $\Psi$. 

\subsection{$L$-functions of Hecke characters}

Associated to a Hecke character $\Psi$ with modulus $\mfrak$ 
 of a number field $K$ with a ring of integers $\cO_K$ is an $L$-function, 
 denoted by $L(\Psi,s)$, which is defined for all ideals $\nfrak \in \cI_\mfrak(\cO_K)$ prime to the 
modulus $\mfrak$ as 
 \beq
  L(\Psi,s) ~=~ \sum_{\nfrak \in \cI_\mfrak(\cO_K)} \frac{\Psi(\nfrak)}{\rmN \nfrak^s},
 \eeq
 where $\rmN\nfrak$ is the norm of the ideal. 
Part of the motivation for introducing these objects is the fact, established by Hecke, that like 
in the case of Dirichlet characters the $L$-functions of Hecke characters admit a product 
  formulation
 \beq
 L(\Psi,s) ~=~ \prod_{\pfrak \in \cI_\mfrak(\cO_K)} \left(1- \Psi(\pfrak) \rmN\pfrak^{-s}\right)^{-1}
 \eeq
 
For modular forms with complex multiplication in the sense of Ribet \cite{r77} there exist algebraic Hecke 
characters $\Psi_f$ such that the $L$-function of the modular form $f$ agrees with that of the character 
 $L(f,s) = L(\Psi_f,s)$.

\subsection{Hodge type of the motives $M(\Psi)$}

Associated to Hecke characters are motives $M(\Psi)$ with complex multiplication such that their 
 $L$-functions agree $L(M(\Psi),s) ~=~ L(\Psi,s)$, where the rhs is the Hecke $L$-function associated
to $\Psi$.  This construction is usually formulated in the context of motives associated to 
abelian varieties, enhanced to a set (category) of motives by including Artin motives 
(see e.g Deligne \cite{d79}).
In the present case the inversion is of more interest, i.e. the recovery of algebraic Hecke characters from 
the $\Om$-motives. This will be described further below for algebraic Hecke characters constructed 
from Gauss sum products.
 
In the case of CM motives the Hodge type $\oplus H^{r,s}(M(\Psi))$ can be made explicit because 
 it is determined by the infinity type 
 of the character $\Psi$. If the field $K$ has degree $\rmdeg~K$ the infinity type is parametrized by 
 integers $S \cong (n_1,...,n_{\rmdeg~K})$. The degree of a CM field is even, hence the integers 
 $n^i$ can be arranged as $(n_1,...,n_{\frac{1}{2}\rmdeg~K}, n_{c\frac{1}{2}\rmdeg~K}, ..., n_{c1})$,
 where $n_{c\ell}$ is associated to the complex conjugate $\osi_\ell$ in the infinity type
 The weight of the infinity type is defined as 
 \beq
  w ~=~ n_\ell + n_{c\ell},
 \eeq
 and it is assumed to be independent of $\ell$. The embeddings $\tau^m$  
 of the coefficient field $E$  act on $S$, leading to transformed infinity types 
 \beq
 \tau^m\circ S ~=~ (n_1^m,...,n_{\frac{1}{2}\rmdeg~K}^m, n_{c\frac{1}{2}\rmdeg~K}^m, ..., n_{c1}^m),
  ~~~~m=1,...,\rmdeg~E.
 \eeq
The Hodge type of the associated motive is given by
 \beq
   H^{n^m_\ell, n_\ell^{cm}}(M(\Psi)) ~=~ H^{n_\ell^m, w-n_\ell^m}(M(\Psi)).
 \eeq
 Over $\mathQ$ the motive is given by
 \beq
 H_{\si_\ell}(M(\Psi)) ~=~ \bigoplus_{m=1}^{\rmdeg~E} ~H^{n_\ell^m, w-n_\ell^m}(M(\Psi)).
 \eeq 
This general picture of algebraic Hecke characters and their associated motives arises in the 
context of weighted hypersurfaces via characters that are induced by the finite field Jacobi sums.
The resulting characters are particularly simple in those cases where the Jacobi sums lead to 
characters associated to imaginary quadratic fields because in this special case it was shown already 
by Hecke that the associated $L$-functions are modular.

\subsection{Jacobi sum Hecke characters}

For Hecke characters of Jacobi sum type the first step is to view the Jacobi sums $j_p(\a)$ 
 associated to finite fields $\mathF_p$ with 
values in a cyclotomic field $K_d = \mathQ(\mu_d)$ as characters 
 on the ideals of this field, leading to cyclotomic characters $J_a(\pfrak)$ 
  defined on prime ideal $\pfrak|p$ in terms of the finite field 
Jacobi sums $j_p(\a)$
 \beq
  J_a(\pfrak) ~~\lolra ~~ j_p(\a), ~~~~a=d\a.
 \eeq
 Here $\a\in \mathQ^{n+1}$ is related to the vectors $u \in \cU_{n,d}$ as 
 $\a_i ~=~ \frac{w_iu_i}{d}$. 
  For the vector $u\in \cU_\Om$ and its Galois images $a_r = \si_r(u_\Om)$ with 
components $a_r^i$ one thus obtains algebraic Hecke characters 
 $\Psi_r := J_{a_r}$ whose structure encoded in 
 the characters
  \beq
  \chi^r_\rmalg(z) ~=~ z^{S_r} ~=~ 
     \prod_{\si_\ell \in \rmGal(K_X/\mathQ)}  \si_{-\ell}^{-1}(z)^{n_r^\ell}
 \lleq{weil-infinity-type-a}
 has been determined by Weil \cite{w49} as
 \beq
  n_r^\ell ~=~ \left[\sum_{j=1}^{n+1} \left \langle \frac{\ell a_r^j}{d}\right\rangle\right],
  \lleq{weil-infinity-type-b}
 where $n+2$ is the rank of the Jacobi sum, $[\cdot]$ is the integral part, and 
 $\langle x\rangle = x -[x]$ is the rational part.

This explicit form of the infinity type for Jacobi sum Hecke characters provides one way to correlate the 
$u$-vectors to their Hodge type.

\section{Algebraic cuspidal automorphic forms and representations}
 
Conformal invariance of string theory makes it reasonable to ask whether the modular forms 
expected to appear in the intermediate cohomology of the compact varieties can be derived from 
the modular forms that arise on the worldsheet. This question leads to a modular realization of the 
notion of an emergent spacetime in string theory \cite{rs08}. Modularity is too narrow a framework, 
however. While all elliptic curves are modular and modular motives also occur in higher dimensions, 
most motives are not modular, necessitating the generalization to automorphic forms and representations.
While all pure motives are expected to be automorphic, not all automorphic forms are 
thought to be motivic. The class of forms that are believed to admit a motivic interpretation is comprised
 of those automorphic objects that are algebraic.
 
\subsection{Automorphic forms}

The focus in this paper will be on automorphic objects associated to $\rmGL(n)$, leading to the 
standard $L$-functions, which are conjectured by Langlands to account for all $L$-functions. While 
several instances of such relations have been established, the general proof of functoriality seems out of reach
at present. The general  notion of an automorphic form is motivated to a large extent by the idea to 
 generalize the group theoretic lift of classical modular forms to $\rmGL(2)$ to higher rank groups.
 This provides a more conceptual view already for modular forms and places these objects into a more 
coherent framework.  Background discussions of automorphic forms and representations can be found in 
 \cite{bj79}.

 Roughly, automorphic forms are complex valued functions $\phi$ on a group $G$  
 that satisfy a number of constraints. First, the covariance of modular forms with respect to some 
 congruence subgroup $\G \subset \rmSL(2,\mathZ)$ is traded for the invariance of $\phi$  with 
 respect to an arithmetic subgroup $\G\subset G(\mathQ)$. Next, the invariance of GL(2) 
 forms with respect to the maximal compact group $K=\rmSO(2,\mathR) \subset \rmSL(2,\mathR)$
 is generalized to $C$-finiteness for a maximal compact subgroup $C(\mathR) \subset G(\mathR)$, 
 i.e. the right translations with respect to $C$ given by 
 $\{\phi(gk) ~{\Big |}~ g\in G, ~k\in C\}$ span a finite 
 dimensional space. The penultimate constraint requires that $\phi$ is annihilated by an ideal in the 
center $Z(U(\gfrak))$ of the universal enveloping  algebra $U(\gfrak)$ of the Lie algebra $\gfrak$ 
 of $G$, and finally it is assumed that $\phi$ is of bounded growth. This structure is made more 
explicit for the GL(2)-automorphic lift of holomorphic modular forms in Section 7 and examples are 
discussed in Sections 9 and 10.

  Associated to automorphic forms are representations $\pi$, induced by the right regular representation 
acting on the space of  square integrable function $L^2(G)$,  that are related to $\phi$ by the fact that 
 the form $\phi$ is an element in an invariant subspace associated to $\pi$.
The key that turns automorphic representations and forms into managable objects is the fact that they 
are characterized by conjugacy classes in the dual group $\Gwhat$. The simplification that appears in 
the context of $\rmGL(n)$ automorphic objects of interest in the present paper is that the dual group 
is again $\rmGL(n)$ and that  it is not necessary the problem of L-packets.

\subsection{Local factors $\phi_v$ and $\pi_v$ and their Langlands parameters}

Similarly to the factorization of the motivic $L$-functions the $L$-functions of Hecke eigenforms 
 decompose into local factors.  The analogous localization of  automorphic structures is given by 
 the tensor product $\pi = \otimes_v \pi_v$ for the automorphic representation, and the corresponding 
 decomposition $\phi=\otimes_v \phi_v$ for the automorphic forms.
 Here $v$ runs through the finite primes $p$ as well as 
the so-called infinite primes
 $v|\infty$ asscociated with the 
 archimedean fields $\mathR$ and $\mathC$. 

In general the local Langlands parameters associated to 
 $\pi_v$ or $\phi_v$ are (equivalence classes of) maps from the Weil-Deligne group $\rmWD(K_v)$ 
 of the local field $K_v$  to the L-group $^LG$, which has as a factor the dual group $\Gwhat$
 \beq
  r_v:~ \rmWD(K_v) ~~\lra ~~ ^LG.
 \eeq
  These maps are of quite different structure, depending on whether $v$ is a finite prime or whether 
 $v|\infty$. 

\subsection{Infinity type of $\pi$}

The conjectural relation between pure motives and automorphic forms and representations 
 becomes most transparent when expressed in terms of the infinity type of the algebraic automorphic 
representations. This structure was  first emphasized in \cite{c90}, which 
can serve as a reference for a more detailed discussion of these objects. 
  For the archimedean case the Weil-Deligne group is just the Weil group $W_{K_v}$  
 and $^LG=\Gwhat$. According to Langlands \cite{rl73-89} the
    archimedean components $\pi_v$, $v|\infty$, 
 are determined by the representations $r_v$ of the Weil groups $W_{K_v}$
  \beq
  r_v: ~~W_{K_v} ~~\lra ~~\Gwhat ~=~\rmGL(r,\mathC)
 \eeq
 if $G$ is the general linear group.
In both cases $v \cong \mathR,\mathC$ the component $\pi_v$ is 
 obtained from a representation of $\mathC^\times \subset W_v$, $v|\infty$, where 
 $\mathC^\times = W_\mathC$. Denoting the restriction of $r_v$ to $\mathC^\times$ by 
  \beq
    r_v|_{\mathC^\times} ~=~ r_\infty
 \eeq
 leads to the map
 \beq
   r_\infty(\pi):~~W_\mathC=\mathC^\times ~\lra ~ \rmGL(r,\mathC)
 \eeq
 which characterizes the archimedean component $\pi_\infty$.
 Since $\mathC^\times$ is an abelian group all  its representations decompose
  into 1-dimensional representations 
 \beq
  r_\infty(\pi) ~=~ \bigoplus_{i=1}^r \chi_\infty^i,
 \eeq
 where the precise form of the characters $\chi_\infty^i$ depends on the structure of $\pi$.

For algebraic automorphic representations the $\chi_\infty^i$  take the form
 \beq
  \chi_\infty^i(z) ~=~ z^{r_i}\oz^{s_i}, ~~~r_i,s_i \in \mathZ
 \eeq
 in an appropriate normalization of $\pi$. (In the literature several different conventions are 
adopted, depending on whether the viewpoint is more motivically oriented or more number theoretically
 oriented. This leads to different conditions for the exponents $r_i,s_i$.)
 The collection of pairs of integers
 \beq
  I_\infty(\pi) = \{(r_i,s_i)~|~ r_i,s_i \in \mathZ\}
 \eeq
  is the infinity type of the algebraic automorphic representation $\pi$.

\subsection{The tensor product of automorphic representations}

One of the main problems in the theory of automorphic motives is to establish the automorphic 
analog of the tensor product. For this purpose automorphic product was introduced by 
Langlands \cite{rl79} some time ago in the context of isobaric representations. 
While the automorphy of this product has not been established in general, for certain classes 
of forms and representations proofs have been found. The most important of these results for the 
present paper is the result of Ramakrishnan for the product $\rmGL(2)\times \rmGL(2)$ 
 \cite{dr00}.

\subsection{Automorphic $L$-functions}

In the case of modular motives $M$ the relation between the spacetime geometry  and the worldsheet 
 theory $T_\Si$ can be made functionally by relating the $L$-function $L(M,s)$ of modular motives 
to  the $L$-function $L(f,s)$ of a modular form $f$ construction from $T_\Si$ 
 \beq
  L(M,s) ~=~ L(f,s).
 \eeq
 The notion of a modular $L$-function is immediate by considering its Fourier 
expansion $f(q) =\sum_n a_nq^n$ and the inverse Mellin transform. Such relations have been 
established in the context of Gepner models and singular fibers of deformations in these models 
in refs. \cite{rs08,kls10}. 
The extension of these results to automorphic motives involves the notion 
of an automorphic $L$-function $L(\pi,s)=L(\phi,s)$, leading to an expected relation of the form 
 \beq
  L(M,s) ~=~ L(\pi,s).
 \eeq
 The precise form of this relation depends on the conventions.

The definition of $L(\pi,s)$  for a representation $\pi$ is most transparent 
for irreducible admissible representations $\pi$ of $\rmGL(n)$  because it is known that each such 
 representation can be factored into a restricted tensor product $\pi = \otimes_v \pi_v$ such that 
 each $\pi_v$ is an irreducible admissible representation of $\rmGL(n)$ over the local fields 
 $K_v$ \cite{f79}.  Furthermore, for almost all $v$ the components $\pi_v$ belong to the unramified 
 principal series representations, 
which give rise to semisimple conjugacy classes $A_v(\pi)$ in $\rmGL(n,\mathC)$ 
 (ref. \cite{ag91} is useful for background material  of these concepts).
 In  general the conjugacy classes belong to the dual group $\Gwhat$ of $G$.
 The eigenvalues of these matrices are often called Satake parameters, and by abuse of notation 
 the matrices $A_v(\pi)$ will be called Satake matrices. 
 The precise form of $A_v(\pi)$ depends on the normalization of 
  the automorphic representation and  the different conventions adopted in the 
  literature result in shifts of the argument $s$ of the $L$-function.
 Given the Satake matrices $A_v(\pi)$, the local factors of the  
 $L$-function of the automorphic representation are  defined as
 \beq
 \cP_v(\pi,t) ~=~ \rmdet\left({\bf 1} - A_vt\right), 
 \eeq
 leading to the global $L$-function 
 \beq
  L(\pi, s) ~:=~ \prod_v \frac{1}{\cP_v(A_v(\pi),p^{-s})}.
 \eeq 
 For a representation $r$ of $\Gwhat=\rmGL(n,\mathC)$ the $L$-function associated to $(\pi,r)$ can be 
defined via $\cP_v(\pi,r,t) ~=~ \rmdet({\bf 1} - r(A_v)t)$.

An agreement of the $L$-function is thus obtained by an agreement of the Hecke 
eigenvalue matrix $H_p(M)$ of the motive and the automorphic Satake matrix $A_v$.

\section{The special cases $\rmGL(2)$ and $\rmGL(4)$}

In the present paper the focus will be on automorphic objects of rank two and four.

\subsection{GL(2) automorphic objects: the space $\rmAutF_\rmhol(\rmGL(2))$}

The structure of automorphic objects can be made most explicit when they are obtained via lifts 
of modular forms. This case in particular allows to make transparent the implications
  of the different 
normalizations that exist in the literature. Since in the present paper the focus is on automorphic motives
it is natural to choose a normalization of the automorphic representations that reflect this relation in the 
simplest possible way. In a later section examples of GL(2)  string automorphic motives arising 
 from nondiagonal Calabi-Yau varieties are discussed in some detail.

 For modular motives the most interesting class of modular forms is that of holomorphic 
cusp forms of some level $N$ with respect to the Hecke congruence group $\G_0$ 
 and some character $\e_N$, the space of which will be
 denoted by $S_w(\G_0(N),\e_N)$. For fixed 
$N$ and weight $w$ this space 
is finite dimensional and formulae can be found in 
 Shimura \cite{gs71}. The lift of a modular form $f \in S_w(\G_0(N),\e_N)$ to an automorphic 
form on $\rmGL(2)$ is obtained by noting that a quotient of the group $\rmGL(2)^+$ generates the 
 upper half-plane $\cH$. Define the action of $g\in \rmGL(2,\mathR)^+$ on $i = \sqrt{-1}$ by 
 \beq
  gi ~=~ \frac{ai+b}{ci+d}, ~~~~g=\left(\matrix{a &b\cr c &d\cr}\right).
 \eeq
 Since the stabilizer of $i$ is given by the subgroup $\rmSO(2,\mathR)$ and the center $Z(\mathR)$, 
 the upper half-plane can  be interpreted as 
 \beq
  \cH ~=~ \rmGL(2,\mathR)^+/Z(\mathR)\rmSO(2,\mathR).
 \eeq

The lift $\phi_f$ of a modular form of weight $w$ is defined as
 \beq
 \phi_f(g) ~=~ \frac{(\rmdeg(g))^{w/2}}{(ci + d)^w} ~f(g(i)), 
 ~~~~~~~~~g=\left(\matrix{a &b \cr c &d\cr} \right) \in \rmGL(2)^+.
 \eeq
 Since $f(\g \tau) ~=~ \e_N(d)(c\tau+d)^wf(\tau)$ for $\g \in \G_0(N) \subset \rmSL(2,\mathC)$ 
 the automorphic  form transforms just with the character $\e_N$ under the congruence group.
This relation can be inverted to construct a modular form $f$ from an automorphic forms $\phi$ as 
 \beq
  f(\tau) ~=~ f(g(i)) ~=~ \left(\frac{ci+d}{\sqrt{\rmdeg~g}}\right)^w \phi(g).
 \eeq
 It is possible to intrinsically identify  within $\rmAutF(\rmGL(2))$ the subspace of 
 cusp forms which is the image of the above lift map (see e.g. ref. \cite{sk04}).

The infinite component $\pi_v$ for $v|\infty$ of GL(2) forms is of discrete series 
 representation type, denoted by $D_w$ for integral $w$, 
 with  a Langlands parameter $r_\infty = r_\infty(D_w) =r_v(D_w)|_{\mathC^\times}$, $v|\infty$, given 
for $\pi=\pi^f$ with $f$ of weight $w$ by
 \beq
  r_\infty(z) ~=~ \left(\matrix{z^{w-1} &0 \cr 0 &\oz^{w-1}\cr}\right).
 \eeq

The local factors $L_p(\pi,s)$ of the $L$-function $L(\pi,s)$ of the automorphic 
 representation $\pi=\otimes_v \pi_v$ at the finite primes $v=p$ are described by rank two 
 Satake matrices $A_p(\pi)$
  whose elements are determined by the Fourier coefficients of the modular form. If $f$ is a Hecke 
 eigenform its Fourier series can be written as an Euler product 
 \beq
  L(f,s)~=~ \prod_p \frac{1}{1-a_p p^{-s} + \e_N(p) p^{w-1} p^{-2s}}.
 \eeq
  Factorizing the polynomial $\cP_p(f,t) ~=~ 1-a_pt +\e_Np^w t^2 = (1-\g_p^{(1)}t)(1-\g_p^{(2)}t)$
 leads to the Hecke matrix 
 \beq
  H_p(f) ~=~ \left(\matrix{\g_p^{(1)}   &0  \cr
                 0  &\g_p^{(2)} \cr} \right) 
 \eeq
 in terms of which the $L$-function can be determined via the local factors 
$\cP_p(f,t) ~=~ \rmdet\left({\bf 1} +H_p(f)t\right)$.

  Identifying the Satake matrix $A_p(\pi)$ with the Hecke matrix 
 \beq
  A_p(\pi) ~=~ H_p(f)
  \eeq
 thus leads to the automorphic $L$-function as
 \beq
  L(\pi,s) ~=~ L(f,s).
 \eeq
 As noted earlier, other normalizations are in use as well, resulting in shifts in $s$.

\subsection{GL(4) automorphic objects}

Automorphic forms and representations associated to higher rank groups behave quite differently from 
 the automorphic forms associated to GL(2), in particular the subspace generated by holomorphic modular 
 forms.  The key here is that these forms are in general not induced by GL(2) forms, even the algebraic 
  automorphic objects that are conjectured to arise from motivic. 
 Part of the difference is that the archimedean components 
 $\pi_\infty$ of $\rmGL(n)$ automorphic forms are not
  of discrete series type for $n>2$,  but instead are tempered. 
 Nevertheless, the infinity type takes a form that is of similar structure as the GL(2) infinity type,
 except that it involves more characters
 \beq
   r_\infty(\pi) ~=~ \rmdiag(\chi_\infty^1, \dots, \chi_\infty^4),
 \eeq
 with 
 \beq
 \chi_\infty^i(z) ~= ~ z^{r_i}\oz^{s_i} ~~~~~r_i,s_i \in \mathZ
 \eeq
 where $z\in \mathC^\times$. 

The question thus becomes how for motivically induced automorphic forms and representations the infinity 
 type is determined by the motive. In the applications below it will become clear that the tempered 
representations of the archimedean components $\pi_\infty$ of $\rmGL(n)$ can form a tensor product 
 of the discrete series type representations $\pi_\infty^{(i)}$ of GL(2) building blocks.

\section{Automorphic motives}

It is generally expected that pure motives are automorphic and that the class of automorphic objects 
obtained in this way coincides with those that are algebraic.

\subsection{Hodge type and infinity type}
The discussion of the cohomology type of the motives and the infinity type of algebraic automorphic 
forms has been summarized above in a form that emphasizes the similarity in structure: both 
types of objects can be represented by vector spaces that admit a filtration characterized by 
 a vector of pairs of integers $(r_i,s_i) \in \mathZ^2$.  If the rank $r$ of the automorphic representations
 is identified with the rank $\rmrk(M)$ of the motive the above description makes it plausible to expect 
a relation of the form
 \beq
 H(M)   ~\cong ~  r_\infty(\pi) \otimes \chi,
 \eeq
 where $\chi$ is a character that takes into account possible twists that may be chosen to implement 
different normalizations of the automorphic representation. If the infinity type $I_\infty(\pi)$ is chosen 
 to be given by the Langlands normalization then the character $\chi$ of a $\rmGL(n)$ representation 
 is usually chosen to take the form $\chi=|\cdot|_\mathC^{(1-n)/2}$ (see e.g. \cite{c90}).

The decomposition of both the cohomology and the 
 infinity type of the algebraic representation $\pi$  leads to the 
more precise relations
 \beq
  \bigoplus_{\stackrel{i=1}{r_i+s_i=w_M}}^{\rmrk(M)} H^{r_i,s_i}(M) 
  ~~\cong ~~ \bigoplus_{i=1}^{\rmrk(M)} \chi_\infty^i\otimes \chi.
\eeq
 
The problem now is to put this conjectural relation into a computable form that can be tested. 
 In the following this will be considered in the context of $\Om$-motives, in particular for 
 weighted Fermat hypersurfaces. 

\subsection{Structure of GL(2) automorphic motives}

In this section the structure of GL(2) automorphic motives is made explicit. The key here is that 
the Hodge structure of the motive can take quite different values because the transition from the 
motive to the automorphic form involves the Tate twist. In the simplest case, relevant e.g. for Calabi-Yau 
varieties, the Hodge structure of an automorphic GL(2)-form of weight $w_\pi = w_f = w+1$ e.g. is 
given by 
 \beq
 H(M) ~\cong ~ H^{w,0}(M) \oplus H^{0,w}(M).
 \eeq
 In general this is not the case however. In the more general class of special Fano varieties, 
 introduced in the context of mirror symmetry for rigid Calabi-Yau varieties 
 \cite{rs92,cdp93,rs94}, the Hodge structure  of the modular motives 
 (associated to GL(2)  automorphic forms) takes the form 
 \beq
  H(M) ~\cong ~ H^{w+Q-1,Q-1} \oplus H^{Q-1,w+Q-1},
 \eeq
 where $Q$ is a positive integer which can be thought of either as the total charge of the underlying 
 Landau-Ginzburg model, or as the codimension of the critical variety associated to this Fano variety.
 More details about these spaces can be found in the above references.

Modular motives of such varieties and their mirror Calabi-Yau varieties have been discussed in 
 \cite{kls08}.  The $L$-function that turns out to be relevant for the associated modular form is not determined 
 by the motivic cardinalities $N_p(M_\Om)$ per se,  but by the renormalized cardinalities defined by 
 \beq
 N_p^Q(M) ~:=~ \frac{N_p(M)}{p^{Q-1}}. 
 \eeq
 The relation between the weight $w_\phi$ of the automorphic form and the weight $w_M$ of motive is 
 \beq
  w_\phi ~= w_M + 1 - 2(Q-1)
 \eeq
  and the infinity type $r_\infty(\pi^Q)$ of the automorphic form of the modular motive takes the form 
 \beq
  r_\infty(\pi^Q) ~=~ \left(\matrix{z^{w_M-2(Q-1)} &0 \cr 0 &\oz^{w_M-2(Q-1)}}\right).
 \eeq

{\bf Remark.} \hfill \break
 Historically, the construction of modular motives has focused on Kuga-Sato varieties, following the 
original work of Deligne \cite{d68}, completed later by Scholl \cite{as90}. In string theory the focus is 
mostly on Calabi-Yau varieties, generalized to the class of special type Fano varieties in the context 
of mirror symmetry for rigid Calabi-Yau spaces.

\section{Automorphic GL(2) motives in dimension two}

In the case of rank two motives the automorphic structure can be linked to the more familiar framework 
of modular forms. This is useful not only because modular motives do appear in higher dimensional 
varieties, but because GL(2) automorphic representations also appear as building blocks of higher 
rank automorphic motives.

Perhaps the simplest nontrivial example of a nondiagonal modular motive is given by the 
 $\Om$-motive of the K3 surface
 \beq
  X_2^{6\rmD} ~=~ \left\{z_0^6+z_1^6+z_2^3 + z_2z_3^2 ~=~ 0\right\} ~\subset ~
  \mathP_{(1,1,2,2)}.
 \eeq
 This variety corresponds to the exactly solvable Gepner model given by the tensor product 
  \beq
  T^{6\rmD}_\Si ~=~ (4_A^{\otimes 2}\otimes 4_D)_\rmGSO, 
  \eeq
 with central charge $c=6$, and the subscript indicates the GSO projection. 
 The Galois group $\rmGal(K_X/\mathQ)$ is determined by $v=\rmlcm \{d_i\}_{i\neq n}=6$,
 hence leads to  
  $\rmGal(\mathQ(\mu_6)/\mathQ) = \{\si_1,\si_5\}$. The $\Om$-motive of this surface is 
  spanned by the Galois orbit of the vector $u_\Om$,  leading to the realization 
 \beq
    H(M_\Om) ~\cong~ \langle \rmGal(K_X/\mathQ), u_\Om\rangle 
  ~=~ u_\Om \oplus \ou_\Om  ~=~ (1,1,1,1)\oplus (5,5, 0,1)
 \eeq
 of the motive, where $\ou_\Om$ is the vector dual to $u_\Om$.

Associated to the motive are Gauss sum products $\mathG_p(u)$ and 
 $\mathG_p(\ou_\Om)$
 which can be used to define the Hecke matrix 
 \beq
  H_p ~=~ -\frac{1}{p}\left(\matrix{\mathG_p(u_\Om) &0\cr 0 &\mathG_p(\ou_\Om)\cr}\right)
 \eeq
 and the local factors $\cP_p(t) ~=~ \rmdet({\bf 1}+H_p(M)t)$.  Here the Gauss sum product 
 $\mathG_p(\ou_\Om)$ for the dual vector $\ou_\Om$ is the complex conjugate of $\mathG_p(u_\Om)$.
 The motivic cardinalities $N_p(M_\Om)$ lead to   
  \beq
   \rmtr ~H_p(M_\Om) ~=~ -N_p(M_\Om),
 \eeq
 and the coefficients of the associated $L$-function are given by $a_p(M_\Om) = \rmtr~N_p(M_\Om)$,
 as described in Section 4. The computation of the Gauss sum products $\mathG_p(u)$ leads to the 
 $L$-function of the $\Om$-motive
  \beq
  L_\Om(X_2^{6\rmD},s) 
      ~\doteq~ 1 - \frac{13}{7^s} - \frac{1}{13^s} + \frac{11}{19^s}  - \frac{46}{31^s} + \frac{47}{37^s} 
   + \cdots
 \eeq  
 where $\doteq$ denotes the incomplete $L$-function as usual.
 The question now is whether this $L$-function is modular and if so, what the associated form is.

One way to establish modularity and to explicitly determine the structure of the resulting modular form 
 is by applying a motivic isomorphism between varieties of $D$-type and varieties of diagonal type 
 \cite{rs13}.  The diagonal varieties will be called to be of $A$-type and denoted by $X^A$, 
  in reference to their affine $A$-type 
  partition function invariants of the underlying exactly solvable conformal field theory. 
  In the present case this motivic $AD$-isomorphism 
 implies that the $\Om$-motive $M_\Om(X_2^{6\rmD})$ 
 of the $D$-type K3 surface is isomorphic to the $\Om$-motive to the diagonal K3 surface given by 
 \beq
  X_2^{6_1\rmA} ~=~ \left\{z_0^6+z_1^6+z_2^6+z_3^2 ~=~ 0\right\}
               ~\subset ~ \mathP_{(1,1,1,3)}
 \lleq{diag-surface6A}
 where the superscript superscript $6_1$ is used because further below a second diagonal degree 6 K3 
surface will appear.

The $\Om$-motive of the surface (\ref{diag-surface6A}) has been analyzed in detail in \cite{rs06}, where it was  shown that its $L$-function is modular in terms of a form $f_{3,27}$ of weight three and level 27 
   \beq
  L_\Om(X_2^{6_1\rmA}, s) ~=~ L(f_{3,27},s).
 \eeq
 The form $f_{3,27}$ is given in closed form by
  \beq
      f_{3,27}(q) ~:=~ \eta^2(q^3)\eta^2(q^9) \vartheta(q^3),
 \eeq
 where $\vartheta(q)$ is a theta series  
 $\vartheta(q) = \sum_{z\in \cO_K} q^{\rmN z}$ associated to the imaginary quadratic Eisenstein 
field $K=\mathQ(\sqrt{-3})$ with $\cO_K$ the ring of integers in this field. 
The string theoretic interpretation of $f_{3,27}$ is most transparent by noting that it is 
 the symmetric square of its $\eta$-product factor, which defines a modular form of weight two and 
level 27. This factor 
 can be expressed in terms of the Hecke indefinite modular forms $\Theta^k_{\ell,m}$ as
 \beq
  f_{2,27}(\tau) ~=~ \Theta^1_{1,1}(3\tau) \Theta^1_{1,1}(q^3), 
 \eeq
 a form that is associated to the elliptic diagonal curve $E^3 \subset \mathP_2$.
The weight one forms 
 \beq
   \Theta^k_{\ell,m}(\tau) ~=~ \eta^3(\tau) c^k_{\ell,m}(\tau)
  \eeq
are obtained in terms of the string functions $c^k_{\ell,m}(\tau)$ of Kac-Peterson 
 \cite{kp84}, associated to the  algebra $A_1^{(1)}$. This affine Lie algebra is the basic building block of 
the $\cN=2$ superconformal models of the Gepner models, hence provides a direct link between the 
worldsheet theory $T_\Si$ and the motive of the compact variety. More details about this diagonal model 
can be found in \cite{rs06}.

The motivic $AD$-isomorphism implies that the $L$-function of the $\Om$-motive of the 
 nondiagonal K3 surface $X_2^6D$ is equal to the $L$-function of the diagonal surface
 \beq
  L_\Om(X_2^{6\rmD},s ) ~=~ L_\Om(X_2^{6_1\rmA},s) 
 \eeq
 hence modular in terms of $f_{3,27}$.
 With this result the automorphic form $\phi$ is given by the lift
  $\phi_{3,27} =\phi_{f_{3,27}}$  and with the appropriate normalization
 the identity $L_\Om(X_2^{6\rmD},s )~=~ L(f_{3,27},s)$
 translates into an identity between the motivic $L$-function and the automorphic $L$-function $L(\phi_{3,27},s)$.

 The motivic $AD$-isomorphism implies furthermore that  the Gauss sum products 
 $\mathG_p(u)$ for $u\in \{u^\Om,\ou^\Om)$ are directly given in terms of the finite field Jacobi sums 
 $j_p(\a_\Om)$ that describe the diagonal surface, where $\a_\Om = \frac{w_i}{d}$.
 This shows that the infinity type of $X_2^{6\rmD}$ can be  computed directly from 
 the Weil formula, leading  
 for $a_\Om= (1,1,1,3)$ and $\oa_\Om=(5,5,5,3)$ to 
  \bea
  S(a_\Om) &=& 2\cdot \si_1 + 0 \si_5 \nn \\
 S(\oa_\Om) &=& 0\si_1 + 2\cdot \si_5.
 \eea
 Thus  the infinity type
 \beq
   r_\infty ~=~ \left(\matrix{\chi_\infty^1 &0 \cr 0 &\chi_\infty^2 \cr}\right)
 \eeq
is given by
     $\chi_\infty^1(z) = z^{S(a_\Om)}$ and $\chi_\infty^2(z) = z^{S(\oa_\Om)}$, leading to 
 the Hodge type of the corresponding motive $M$ of the nondiagonal variety 
 whose weight is given by $w_M = n_1+n_5=2$
 \beq
  H(M) ~=~ H^{2,0}(M) \oplus H^{0,2}(M),
 \eeq
 In the present case the motive is pure and the weight of the motive  is given by the degree of the cohomology.

 The key structure of such GL(2) automorphic representation is that the space of functions is 
 1-dimensional and that the corresponding automorphic form $\phi=\phi_\pi$ is determined by 
the lift of a cusp modular form $f(\tau)$ on the complex upper half-plane of some level $N$ and 
a weight $w_f$ that is determined by the weight $w_M$ of the motive as $w_f=w_M+1$.

The infinite component $\pi_\infty$ of the automorphic representation $\pi= \otimes_v \pi_v$
 only determines the infinite factor of the completed $L$-function of the motive. The main arithmetic 
 information is contained in the local factors $\pi_p$ at the finite primes, similar to the local 
factors $L_p(M_\Om,s)$ of the motive, obtained from the local zeta functions 
 $Z(X/\mathF_p,t)$. Identifying the Satake matrix $A_p(\pi)$ is identified with the Hecke matrix 
 \beq
  A_p(\pi) ~=~ H_p(M_\Om)
 \eeq
 then leads to the above relation between the motive and the automorphic representation determine 
 via $\phi_{3,27}$ as a lift of a weight 3 form with respect to the Hecke congruence group $\G_0(27)$.

Other nondiagonal K3 surfaces of $D$-type or of more general type can be discussed in a similar way.
An example that does not correspond to an exactly solvable model of Gepner type but is string 
modular nevertheless is given by the K3 surface 
 \beq
 X_2^{12\rmND} ~=~ \left\{z_0^6 + z_1^3 + z_2^4 + z_2z_3^3 ~=~ 0\right\}
  ~\subset ~\mathP_{(2,4,3,3)}.
 \eeq
 In this case the field $K_X$ of the variety is given by $\mathQ(\mu_v)$ with (here $n=2$)
 \beq
 v ~=~\rmlcm\{d_i\}_{i\neq n} ~=~ 6,
 \eeq
 hence the rank of the motive is $\rmrk~M_\Om(X_2^{12\rmND})=2$ and therefore is modular 
because the motive is of CM type. The modular form $f_{3,48}$ such that 
 \beq
 L_\Om(X_2^{12\rmND},s) ~=~ L(f_{3,48},s).
 \lleq{modulark3-2}
 can be shown to be of weight three and  level $N=48$, with the closed form expression
\beq
  f_{3,48}(q) = \eta^3(q^2)\eta^3(q^6)\otimes \chi_3,
 \lleq{weight3-level48} 
 where $\chi_3$ is a Legendre character of the type $\chi_n(p) = \left(\frac{n}{p}\right)$:

The form (\ref{weight3-level48}) was shown in \cite{rs06} to be the symmetric 
square of the elliptic modular form $f_{2,144} \in S_2(\G_0(144))$ which has a string interpretation 
in terms of the Hecke indefinite modular form $\Theta^1_{1,1}(\tau)$ as 
 \beq
  f_{2,144}(q) ~=~ \Theta^1_{1,1}(q^6)^2 \otimes \chi_3. 
 \eeq
 This reflects the fact that the K3 surface $X_2^{12\rmND}$ is an elliptic fibration with a generic elliptic 
fiber in the configuration $E^6(\l) \subset \mathP_{(1,2,3)}$. The $L$-function of the diagonal curve 
 $E^6$ was computed in \cite{rs05} and is given by the weight two form above, leading to  
 \beq
 L(E^6,s) ~=~ L(f_{2,144},s).
 \eeq
It would be interesting to find an exactly solvable 
 conformal field theory construction of this K3 surface that explains the 
 $L$-function identity (\ref{modulark3-2}).

\section{An automorphic GL(4)-motive in dimension three}

In this section the automorphic structure of a  rank four motive 
 of the nondiagonal Calabi-Yau threefold
 \beq
  X_3^{12\rmD} ~:=~ \left\{z_0^{12}+z_1^{12}+z_2^6 + z_3^3+z_3z_3^2 ~=~ 0\right\}
   ~\subset ~ \mathP_{(1,1,2,4,4)}
  \eeq
 is considered. This variety is associated to an exactly solvable tensor model of the form 
 \beq
 T^{12\rmD}_\Si ~=~ \left(10_A^{\otimes 2} \otimes 4_A \otimes 6_D\right)_\rmGSO
 \eeq
 on the worldsheet.
 
 The number field associated to this variety is the cyclotomic field 
 $K_X = \mathQ(\mu_{12})$, which leads to the realization of the $\Om$-motive as 
 \bea
  H(M_\Om) 
   &=& \langle \rmGal(K_X/\mathQ) , u_\Om\rangle \nn \\
   &=& (1,1,1,1)\oplus (5,5,5,0,1) \oplus (7,7,1,1,1) \oplus (11,11,5,0,1).
 \eea
 The local factors $L_p(M_\Om,s)$ of the $L$-function of the $\Om$-motive are therefore given by 
the Gauss sum products associated to the motivic vectors $\si_\ell(u_\Om) \in \cU_\Om$
 \beq
  H_p(M_\Om) ~=~ \frac{1}{p}\left(\matrix{\mathG_p(\si_1(u_\Om)) &  &  & \cr
                                                    &\mathG_p(\si_5(u_\Om))  &  & \cr
                                   &   &\mathG_p(\si_7(u_\Om))   & \cr
                   &   &   &\mathG_p(\si_{11}(u_\Om))\cr}\right)
 \eeq
  via the polynomials
 \beq
  \cP_p(M_\Om,t) ~=~ \rmdet\left({\bf 1}  +  H_p(M_\Om)t\right) 
    ~=~ \prod_{\ell \in (\mathZ/12\mathZ)^\times} \left(1 + \frac{1}{p}\mathG_p(\si_\ell(u_\Om))t\right).
 \eeq
The resulting projective motivic cardinalities lead to the $L$-function coefficients 
 \beq
  a_p(M_\Om) ~=~ - \frac{1}{p} \sum_{u\in \cU_\Om} \mathG_p(u),
 \eeq
 whose computation leads to the $L$-function 
  \beq
 L_\Om(X_3^{12\rmD},s) 
 ~\doteq~ 1 - \frac{132}{13^s} + \frac{52}{37^s} + \frac{740}{61^s} - \frac{276}{73^s}
   - \frac{36}{97^s} - \frac{1284}{109^s} + \cdots 
 \lleq{Lfunction}
 with $\doteq$ indicating the incomplete $L$-function as before.

The question is again whether this $L$-function is automorphic, and if so, what 
its structure is.
The structure of this $L$-function can be understood by proceeding in a way analogous 
 of the strategy adopted in \cite{rs08} for  rank four automorphic motives derived from diagonal varieties.
 The idea is to find modular motivic building blocks and to build the $L$-function of the $\Om$-motive 
of the threefold via the Rankin-Selberg convolution of the lower-weight modular forms. 
  In the present example it can be checked that the $L$-function of the modular form 
 (\ref{weight3-level48}) of weight three and level 48 
 factors into the $L$-function (\ref{Lfunction}). It was shown in \cite{rs06} 
 that this modular form is $\Om$-motivic, with the motive coming from the K3 hypersurface 
 \beq
  X_2^{6_2\rmA} ~=~ \left\{z_0^6+z_1^6+z_2^3+z_3^3 ~=~0\right\} ~\subset ~\mathP_{(1,1,2,2)}.
 \eeq
 As noted above, the string theoretic nature of the 
 modular form (\ref{weight3-level48})  becomes apparently by noting that it is 
 the symmetric square of a modular 
 form of weight two and level 144 which is given in terms of the Hecke indefinite modular
  forms $\Theta^k_{\ell,m}$ as $f_{2,144}(q) ~=~ \Theta^1_{1,1}(q^6) \otimes \chi_3$.

 The quotient series obtained from the two $L$-series $L_\Om(X_3^{12\rmD}$ and 
  $L_\Om(X_2^{6_2\rmA},s)$ is modular of weight two and level 256, hence the 
 threefold $L$-function is the Rankin-Selberg convolution of two modular forms, much like in the 
case of the automorphic rank four motives discussed in ref. \cite{rs08}. 

The proof of the automorphy of the rank four motive of the CY threefold can be completed by 
 applying again the motivic isomorphism of \cite{rs13}  between motives of $D$-varieties and motives 
 of $A$-varieties. In the present case this isomorphism 
implies that the $\Om$-motive of the nondiagonal variety $X_3^{12\rmD}$ is identical 
to  the $\Om$-motive of the 
 diagonal Calabi-Yau threefold 
 \beq
  X_3^{12\rmA}  ~:=~ \left\{z_0^{12} + z_1^{12} + z_2^6+z_3^6 + z_4^2 ~=~ 0\right\} ~
       \subset ~     \mathP_{(1,1,2,2,6)},
  \eeq
 which is known to be automorphic. General discussions aimed at nonexplicit automorphy obtained by 
base change for cyclic extension fields can be found in \cite{rl80, ac89} for extensions of prime degree and for 
 non-prime degree in \cite{lr98}.  Automorphy of the Rankin-Selberg convolution of GL(2)-forms 
 was established in ref.  \cite{dr00}.

 This result allows to compute the infinity type of the algebraic Hecke characters of the nondiagonal 
 CM motive 
  \beq
   S(\si_r(a_\Om)) ~=~ \sum_{\ell \in (\mathZ/d\mathZ)^\times} n_r^\ell \si_\ell
  \eeq
 via the characters $J_{\si_r(a_\Om)}$  associated to $\mathQ(\mu_d)$ with 
 $d=12$. Weil's formulae  (\ref{weil-infinity-type-a}) and (\ref{weil-infinity-type-b})  
  lead to
  \bea
  S(\si_1(a_\Om)) &=& 3\si_1 ~+~ \si_5 + 2 \si_7 + 0\si_{11} \nn \\
  S(\si_5(a_\Om)) &=& \si_1 + 3\si_5 + 0\si_7 + 2\si_{11} \nn \\
 S(\si_7(a_\Om)) &=&  2\si_1 + 0\si_5 + 3\si_7 + \si_{11} \nn \\
  S(\si_{11}(a_\Om)) &=&  0\si_1 ~+ 2\si_5 + \si_7 + 3\si_{11}.
 \eea
 Put slightly different, the infinity type associated to the $\Om$-motive $M_\Om$ can be viewed as 
 an infinity type matrix $S=(n_r^s)$, denoted by the same symbol. 

The infinity matrix $S$ determines the Hodge type of the motive via the motivic weight
  $w_M=n_r^s + n_r^{cs} = 3$ as 
 \beq
  H_{\si_r}(M_\Om) ~=~ \bigoplus_{s=1}^{\rmdeg~E} H^{n_r^s,w-n_r^s}
   ~=~ H^{3,0}\oplus H^{1,2} \oplus H^{2,1} \oplus H^{0,3}.
 \eeq
 
In the present case the automorphy of the diagonal hypersurface implies that modular Hecke matrices 
 coming from the $\Om$-motive of the K3 surface $X_2^{6\rmB}$
  \beq
  H_p(M_\Om(X_2^{6\rmB})) 
     ~=~ \frac{1}{p} \left(\matrix{\mathG_p(u_\Om^{(2)}) &0  \cr  0  &\mathG_p(\ou_\Om^{(2))}\cr}
   \right),
 \eeq
 and the Hecke matrix of the modular form $f_{2,256}$, given by its Fourier coefficients 
  $f_{2,256}(q) = \sum_n a_n(f)q^n$  via 
 \bea
   a_p(f)  &=& \a_p(f) + \b_p(f) \nn \\
    p  &=& \a_p(f)\b_p(f)
 \eea
  as 
 \beq
   H_p(f)~=~ \left(\matrix{\a_p(f) &0 \cr 0 &\b_p(f)\cr }\right)
 \eeq
 lead to the rank four Hecke matrix of the threefold by considering the tensor product
 \beq
  H_p(M_\Om(X_3^{12\rmD})) ~=~ H_p(M_\Om(X_2^{6\rmB})) \otimes H_p(f).
 \eeq

The infinity types of the algebraic Hecke characters then lead to 
the following Langlands parameter of the 
  archimedean component
 \beq
  r_\infty(z)  ~=~ \left(\matrix{ z^3  &   &  & \cr
                                                   &z\oz^2 &  & \cr
                                                  &             &z^2 \oz & \cr
                 &  &   &\oz^3 \cr}\right).
 \eeq

\vskip .2truein 

\section{Outlook} 

In the present paper the focus has been on the automorphic structure of nondiagonal spacetimes 
 whose underlying conformal field theory structure is given by exactly solvable theories  that are 
of nondiagonal Gepner type, or Kazama-Suzuki type. Within these classes of varieties one can 
consider family spaces, which in the underlying conformal field theory correspond to 
 deformations along marginal operators. Such deformations have previously been considered 
 for diagonal Gepner models, with a particular focus on fibers in the families that are modular. 
 For such points in the moduli space it is possible to ask whether the motivic modular forms 
are again related to the modular forms on the worldsheet theory $T_\Si$, hence whether the 
geometric forms admit a string theoretic interpretation. In ref. \cite{kls10}  such string 
theoretic modularity was shown to exist for mixed motives that arise from singular fibers in 
deformation families of diagonal varieties. This shows that the phase transitions between 
topologically distinct Calabi-Yau varieties described by modular mixed motives encode at least 
part of the conformal field theoretic structure of the rational points. This is a strong 
indication that the modular phase transitions that arise in these families are string theoretically 
 consistent. 
 
The results described in the present paper on automorphic motives of nondiagonal rational theories 
suggest a two-fold extension of the family analysis in \cite{kls10}.  It would be 
 interesting to  consider fibers in the deformation families diagonal Gepner models that give 
rise to automorphic forms and representations rather than modular forms and to relate these 
autormorphic structures to the conformal field theory on the worldsheet.
 For smooth fibers this involves automorphic pure motives, but in the case of singular fibers 
describing phase transitions this involves the concrete construction of automorphic representations 
 associated to mixed motives. Such automorphic mixed motives are at present not understood, 
 but conjecturally provide the most general framework that exists at present time.

In the context of the worldsheet theory work remains to done in particular in the context of $G/H$ 
 exactly solvable theories. Missing in these constructions extending the SU(2)/U(1) theory underlying 
the Gepner models is an understanding of the string functions of Kac and Peterson that play a pivotal 
role in the Gepner models. The explicit construction of these objects along the lines of Kac-Peterson
  \cite{kp84}
 would provide an important step in extending the automorphic spacetime program to a more comprehensive 
 framework.

\vskip .3truein

{\bf Acknowledgement.} \hfill \break
 It is a pleasure to thank Monika Lynker for discussions. This research has been supported in part by a 
 grant from the NSF under grant no. NSF-0969875.  The work reported here 
 has been conducted over a long period, 
during which I have benefitted from the support and hospitality of several institutions. 
 Visits to the Werner Heisenberg Max Planck Institute in 2011 and to CERN in in 2012 have greatly 
  facilitated progress on this project and I'm grateful for support from the Max Planck 
  Gesellschaft and CERN. Special thanks are due in particular to Dieter L\"ust in Munich and Wolfgang 
 Lerche at CERN for making these visits possible and I thank the string theory groups at both 
 institutitons for their friendly hospitality.

\end{document}